\documentclass[prl,twocolumn,superscriptaddress,showpacs]{revtex4}
\usepackage{graphicx}

\begin{document}


\title{From ballistic motion to localization: a phase space analysis}
\author{Andr{\'e} Wobst}
\author{Gert-Ludwig Ingold}
\author{Peter H{\"a}nggi}
\affiliation{Institut f{\"u}r Physik, Universit{\"a}t Augsburg, 
Universit{\"a}tsstra{\ss}e 1, D-86135 Augsburg, Germany}
\author{Dietmar Weinmann}
\affiliation{Institut de Physique et Chimie des Mat{\'e}riaux de Strasbourg, 
UMR 7504 (CNRS-ULP), 23 rue du Loess, F-67037 Strasbourg Cedex, France}
\date{1 October 2001}

\begin{abstract}
We introduce phase space concepts to describe quantum states in a 
disordered system. The merits of an inverse participation
ratio defined on the basis of the Husimi function are demonstrated 
by a numerical study of the Anderson model in one, two, and
three dimensions. Contrary to the inverse participation ratios in real 
and momentum space, the corresponding phase space quantity allows for
a distinction between the ballistic, diffusive, and localized regimes
on a unique footing and provides valuable insight into the structure 
of the eigenstates.
\end{abstract}
\maketitle

The behavior of a quantum particle in a disorder potential depends
significantly on the disorder strength. If the mean free path exceeds
the system size, one may think of plane waves characterized by a fixed
momentum which are scattered by the weak random potential. On the other
hand, sufficiently strong disorder leads to exponentially localized
states in real space. Then a wide range of momenta is needed to construct
these states. A description for arbitrary disorder strength thus requires
to adequately take into account real space as well as momentum space
properties. In this paper, we will therefore adopt a phase space approach.

Signatures of the different regimes can already be found in the energy 
spectrum. While for weak and strong disorder energy levels can almost be
degenerate, level repulsion occurs in an intermediate regime. Making use
of random matrix theory the chaotic nature can be verified and related to 
diffusive motion. However, such considerations cover only 
statistical properties of the spectrum and do not give information about the 
structure of individual states.

A popular way to investigate the properties of single states
is the calculation of their inverse participation ratio. In real space this
quantity has frequently been employed \cite{mirli00} to measure the
size of the localization domain of quantum states in the localized
regime and to characterize the Anderson transition
\cite{abrah79,lee85}. However, the real space inverse participation ratio is
not very sensitive to changes in extended wave functions when going
from the ballistic to the diffusive regime. In order to obtain a meaningful 
description of the structure of quantum states for all regimes on a unique 
footing, we generalize the concept of the inverse participation ratio to 
phase space.

The Anderson model of disordered solids \cite{ander58} has been the subject of 
extensive investigations over the last decades \cite{krame93}. A numerical
study of this model in one, two, and three dimensions will demonstrate the
virtues of our approach. In particular, we will be able to identify ballistic,
diffusive, and localized regimes via properties of the eigenstates. In one 
dimension the results will significantly differ from those in higher
dimensions, which can be attributed to the absence of a diffusive regime.

We start by introducing the relevant phase space concepts. At this point there
is no need to specify the details of the disordered system except that we will 
consider a $d$-dimensional lattice model with lattice constant $a$ and length 
$L$ in each direction. In order to keep the notation simple, we will give the 
formulae for the case of one dimension, which can be generalized to higher 
dimensions in a straightforward manner. 

A positive definite density in phase space is given by the Husimi function
\cite{husim40} or Q function \cite{cahil69}
\begin{equation}
\varrho(x_0,k_0)=|\langle x_0, k_0|\psi\rangle|^2\,.
\label{eq:rhodef}
\end{equation}
Here, the state $|\psi\rangle$ is projected onto a minimal uncertainty
state $|x_0,k_0\rangle$ centered around position $x_0$ and momentum $k_0$. 
In position representation, the latter assumes a Gaussian form
\begin{equation}
\langle x|x_0,k_0\rangle=
\left(\frac{1}{2\pi\sigma^2}\right)^{1/4}
\exp\left(-\frac{(x-x_0)^2}{4\sigma^2}
           +i k_0 x\right)
\label{eq:coh}
\end{equation}
where the value of the variance $\sigma^2$ is yet undetermined. The Husimi 
function is normalized,
$\int(dxdk/2\pi)\varrho(x,k)=1$, and, for real wave functions $\psi(x)$, obeys 
the symmetry $\varrho(x_0,k_0) = \varrho(x_0,-k_0)$.

\begin{figure*}
\includegraphics[width=\textwidth]{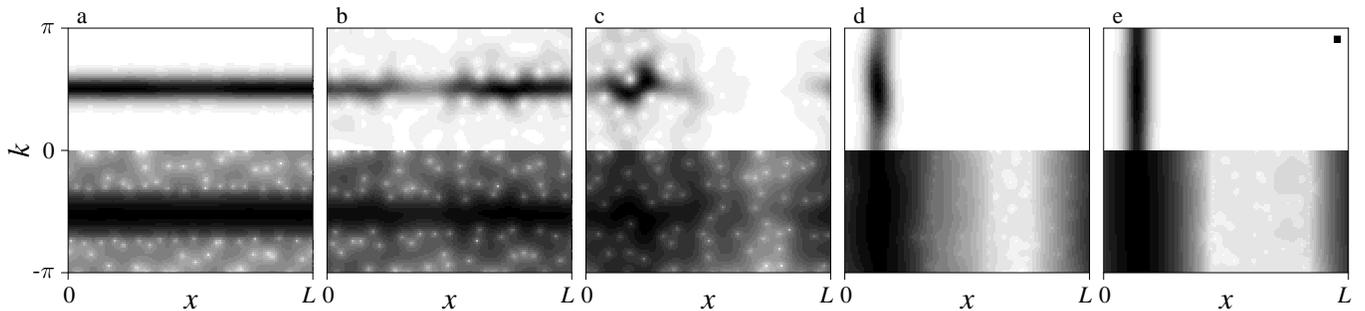}
\caption{The Husimi function of a state at the band center is shown for a 
one-dimensional Anderson model of size $L=128$ with disorder strengths 
$W=0.1, 1, 2.5, 10$, and $25$ increasing from left to right. Exploiting the 
symmetry with respect to $k=0$ the Husimi function is plotted on a linear and 
logarithmic scale in the upper and lower half, respectively. The gray values 
from white to black for increasing Husimi function have been normalized 
separately in each panel. The width $\sigma$ of the Gaussian 
(\protect\ref{eq:coh}) is indicated by the black square in panel (e).}
\label{fig:hd}
\end{figure*}

The variance $\sigma^2$ of the Gaussian (\ref{eq:coh}) determines the relative 
importance of real and momentum space. In the following, 
we choose $\sigma^2=La/4\pi$ leading to equal 
widths of the Gaussian in $x$- and $k$-direction. In order to obtain 
sufficient resolution, one has to ensure that $\sigma\ll L$ which limits the 
possible system sizes from below. Then, the effect of neglecting the tails of 
the Gaussian in finite size systems in presence of periodic boundary condition 
will also be small. Examples of Husimi functions for one-dimensional disordered 
systems are presented in Fig.~\ref{fig:hd}, which will be discussed in detail 
below.

In higher dimensions, Husimi functions can no longer be visualized easily.
A more global description of the phase space properties of the state is
therefore necessary and often sufficient. Based on the Husimi function 
the so-called Wehrl entropy \cite{wehrl79,mirba95,gnutz01} is defined, which 
represents a measure of the phase space occupation. It was shown 
for the driven rotor that the Wehrl entropy of individual quantum states is 
connected to the energy level statistics \cite{gorin97}. A very similar system, 
the kicked rotor, can be mapped onto the Anderson model \cite{fishm82}, 
suggesting that the Wehrl entropy is a useful quantity for the characterization 
of the eigenstates of the Anderson model \cite{weinm99}.

For practical purposes it is more convenient to introduce the inverse
participation ratio in phase space
\begin{equation}
P=\int\frac{dx\,dk}{2\pi}\,[\varrho(\vec x,\vec k)]^2\,,
\label{eq:ipr}
\end{equation}
which corresponds to a linearization of the Wehrl entropy \cite{varga94}.
This inverse participation ratio can be compared directly to the 
corresponding quantities $P_x=\int dx\vert\psi(x)\vert^4$ and 
$P_k=\int dk\vert\tilde\psi(k)\vert^4$ in real and momentum space, 
respectively. $P_x$ is particularly well studied as it is related to 
the probability for a diffusing particle to return to its original
position in the long-time limit \cite{thoul74}.

Furthermore, $P$ can be evaluated without recourse to the $2d$-dimensional
Husimi function. In fact, only the wave function $\psi(x)$ is required 
since we can recast (\ref{eq:ipr}) in the form
\begin{eqnarray}
\label{eq:preal}
P=\frac{1}{8\sqrt{\pi}\sigma}
\int du\,\left|\int dv\,
\psi\!\left(\frac{1}{2}(u-v)\right)
\psi\!\left(\frac{1}{2}(u+v)\right)\right.\\
\times\left.\exp\left(-\frac{v^2}{8\sigma^2}\right)\right\vert^2\,.
\nonumber
\end{eqnarray}
Its nondiagonal character provides the information on momentum. It is only 
by means of (\ref{eq:preal}) that one succeeds in determining the inverse 
participation ratio for three-dimensional systems.

In the following, we specifically consider the Anderson model for 
non-interacting electrons on a lattice with periodic boundary conditions, 
where disorder is modeled by a
random on-site potential. In $d=1$, the Hamiltonian reads
\begin{equation}
H=-t \sum_n (|n\rangle\langle n+1|+|n+1\rangle\langle n|)+
W \sum_n v_n|n\rangle\langle n|
\label{eq:ham}
\end{equation}
with Wannier states $|n\rangle$ localized at sites $n=1, \dots, L$. All 
lengths are measured in units of the lattice constant $a=1$. The hopping
matrix element $t=1$ between neighboring sites defines the energy
scale. The on-site energies $v_n$ are drawn independently from
a box distribution on the interval $[-1/2;1/2]$ and $W$ denotes the
disorder strength.

Husimi functions for a state at the band center are presented in 
Fig.~\ref{fig:hd} for increasing disorder strength $W$ and a randomly selected 
disorder realization $v_n$. Making use of the symmetry 
with respect to $k=0$, the Husimi function is plotted on a linear and 
logarithmic gray scale in the upper and lower half, respectively. White 
points in the lower half may be related to the zeros of the Husimi function
\cite{leboe90}. 

For $W=0.1$ (Fig.~\ref{fig:hd}a), the disorder represents only a small
perturbation and the Husimi function is thus still close to that of plane
waves. Except for the states at the band edges, one finds two stripes
well localized at the corresponding $k$-values, which are extended over
the full real space. The width of the stripes is induced by
the projection onto the Gaussian (\ref{eq:coh}).
In Fig.~\ref{fig:hd}e the opposite limit is depicted. At $W=25$, the
hopping is a small perturbation, $t\ll W$, and the state is localized
in real space. 

In Fig.~\ref{fig:hd}d, where $W=10$, the influence of
the nearest neighbor hopping becomes relevant and tends to extend the
wave function in real space over several sites. Since the coupled
states have to remain orthogonal, they separate in momentum space.
This leads to a contraction of the Husimi function in $k$ direction,
which is more important than the spreading over a few sites. As a
consequence, the phase space properties for strong disorder are
dominated by the behavior in momentum space and the inverse
participation ratio in phase space increases with decreasing disorder.
This behavior will also be evident from Figs.~\ref{fig:ipr1d} and 
\ref{fig:iprnd} below.

Similarly, at weak disorder, the transition from Fig.~\ref{fig:hd}a to
Fig.~\ref{fig:hd}b, i.e.\ to $W=1$, can be understood in terms of a coupling
between different plane waves. Here, however, the coupling is not
restricted to neighboring $k$ values, but is governed by the energy
difference of the respective states. For one-dimensional systems, the
contraction in real space dominates the spreading in
momentum space, again leading to an increase of the inverse
participation ratio in phase space (cf.\ Fig.~\ref{fig:ipr1d}b). 

As a consequence of the behavior for weak and strong disorder, one
expects a maximum for the inverse participation ratio at intermediate
disorder strength. Indeed, for $W=2.5$, the state shown in Fig.~\ref{fig:hd}c 
displays strong localization in phase space.

The situation just described is generic for one-dimensional
systems. This can be seen from the distributions of the inverse
participation ratio depicted in Fig.~\ref{fig:ipr1d}. The distributions
of the logarithms of $P$ as well as $P_x$ (real space) and $P_k$
(momentum space) have been obtained by diagonalizing Eq.~(\ref{eq:ham})
for 50 different disorder realizations $v_n$ for each disorder strength
$W$ and taking $L/2$ states around the band center into account.

\begin{figure}
\includegraphics[width=\columnwidth]{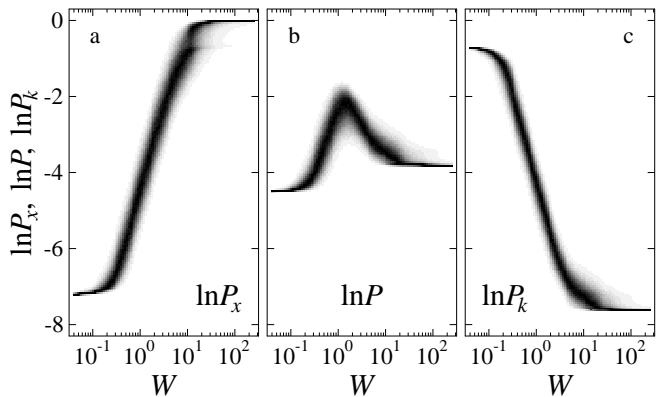}
\caption{Distributions of the logarithms of the inverse participation
ratio for 50 disorder realizations in (a) real space, (b) phase space, 
and (c) momentum space as a function of the disorder strength $W$ for a 
one-dimensional system of length $L=2048$.}
\label{fig:ipr1d}
\end{figure}

In the limit of very strong disorder, the states are localized 
on a single site in real space and uniformly distributed over momentum 
space. This leads to the limiting values $P_x(\infty)=1$ and $P_k(\infty)
=L^{-d}$. In phase space one has to account for the finite width of the
Husimi function and thus finds $P(\infty)=L^{-d/2}$. For $d=1$, these
values can be checked against the data shown in Fig.~\ref{fig:ipr1d}. 
Starting from this limit, with decreasing disorder two energetically almost 
degenerate states become
coupled via the finite hopping matrix element $t$. For these states
$P_x$ is reduced to $1/2$ while $P_k$ is enhanced by a factor of $3/2$. 
As already discussed above, it is the latter which dominates the
behavior in phase space.

In the opposite limit $W\to0$, the real wave functions in $d=1$ contain equal
contributions from degenerate plane waves of momenta $k$ and $-k$. This
implies $P_k(0)=1/2$ and $P_x(0)=2/3L$. In phase space, the finite width
of the Husimi function leads to $P(0)=1/2L^{1/2}$. In higher dimensions,
however, degeneracies occur and render the behavior for $W\to0$ more 
complex. Nevertheless, as a function of system size, $P_x$ 
and $P$ scale as $L^{-d}$ and $L^{-d/2}$, respectively. A more detailed 
discussion will be given elsewhere \cite{wobstxx}.

\begin{figure}
\includegraphics[width=\columnwidth]{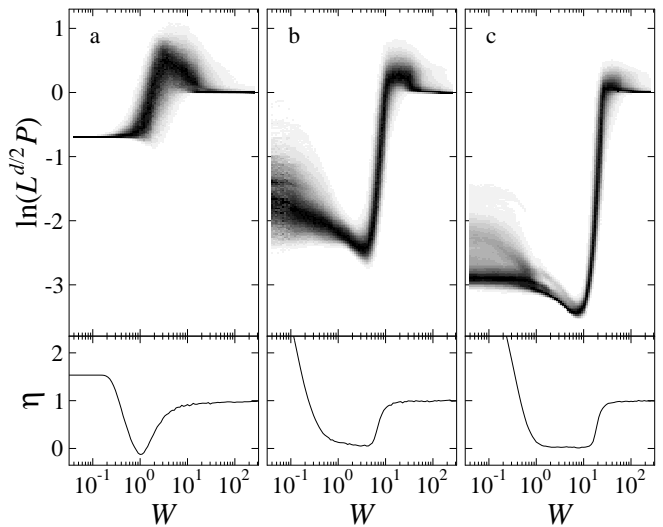}
\caption{Distributions of the logarithms of the inverse participation
ratio in phase space (top) and $\eta$ defined in Eq.~(\protect\ref{eq:eta})
(bottom) are shown as a function of disorder strength $W$ for
(a) $d=1, L=128$, (b) $d=2, L=64$, and (c) $d=3, L=20$. The distributions
are based on 250, 20, and 20 disorder realizations, respectively.}
\label{fig:iprnd}
\end{figure}

The upper part of Fig.~\ref{fig:iprnd} shows the distributions of the
logarithm of $P$ for dimensions $d=1, 2,$ and 3 and system sizes $L=128, 64,$
and 20, respectively. For better comparison the data have been scaled with
the length dependence $L^{-d/2}$, being valid in the limits $W\to0$ and 
$\infty$.

The most striking difference between the one-dimensional case and higher
dimensions consists in the behavior of the inverse participation ratio for weak
disorder. For $d=1$, the average inverse participation ratio increases with
disorder strength and eventually goes through a maximum. The 
overall behavior can be understood as a crossover from a regime dominated
by real space properties to one dominated by momentum space. 

In contrast, in $d\ge2$ the average inverse participation ratio initially
decreases. This implies a spreading of the Husimi function beyond the broadened 
stripes present for weak disorder (cf.\ Fig.~\ref{fig:hd}a). On the other hand, 
the behavior at strong disorder is governed by the same mechanism as
for the one-dimensional case, which implies an increase of the inverse 
participation ratio with decreasing disorder. The two regimes are joined by 
a short interval of disorder strengths where the Husimi function contracts 
strongly as disorder is increased.

The distributions of $P$ depicted in Fig.~\ref{fig:iprnd} behave similarly for
$d=2$ and 3. However, the pertinent scaling argument for Anderson localization 
\cite{abrah79} predicts a phase transition only in dimensions higher than two.
Unfortunately, in $d=3$ numerical constraints prevent us from performing a 
finite size scaling, which would allow to distinguish the form of the jumps of 
$P$ in $d=2$ and 3 in the thermodynamic limit $L\to\infty$.

The main difference between the behavior in $d=1$ and $d\ge2$, the appearance
of a minimum in the inverse participation ratio $P$, can be associated with
the existence of a diffusive regime in $d\ge2$. While in $d=3$ a diffusive 
regime appears even in the
thermodynamic limit, in $d=2$ it is present for systems of finite size
when the system size exceeds the mean free path but not the localization
length. It is suggestive to conclude that diffusive motion is associated with 
a large spread in phase space and thus with the minimum observed in the 
inverse participation ratio. 

To substantiate this idea, we compare our results to energy level statistics.
In the diffusive regime the energy spacing distribution $p(s)$ is close to the 
Wigner-Dyson distribution $p_W(s)=\frac{1}{2}\pi 
s\exp(-\pi s^2/4)$.
In contrast, in the localized regime, $p(s)$ approaches the Poissonian 
statistics $p_P(s)=\exp(-s)$. To quantify the form of the distribution we 
evaluate
\begin{equation}
\eta=\frac{\int_0^bds(p(s)-p_W(s))}{\int_0^bds(p_P(s)-p_W(s))}
\label{eq:eta}
\end{equation}
which is particularly sensitive to level repulsion. Here, $b=0.4729\dots$ 
refers to the first crossing point of the distributions $p_W(s)$ and $p_P(s)$. 
According to its definition, $\eta=1$ for a Poissonian spacing distribution and 
$\eta=0$ for a Wigner-Dyson spacing distribution.

In the lower part of Fig.~\ref{fig:iprnd}, $\eta$ is shown as a function of
the disorder strength. For weak disorder, $\eta$ exceeds 1
because of non-universal level statistics appearing in regular geometries in
the ballistic regime. For strong disorder one finds Poissonian statistics 
as expected for
the localized regime. At intermediate disorder strengths in $d=1$, $\eta$ 
exhibits a minimum due to level repulsion, but no proper diffusive regime
exists. On the other hand, for finite size systems in $d=2$, an extended region 
is present, where the level statistics is close to that predicted by random 
matrix theory. This property is commonly used to identify a region of diffusive 
dynamics. For $d=3$ the diffusive region survives even in the thermodynamic 
limit $L\to\infty$.

Fig.~\ref{fig:iprnd} clearly shows that the decrease of the inverse 
participation ratio $P$ is related to a plateau of $\eta$ at values close to
zero. Using the spreading of the states in phase space as an indicator for 
diffusive behavior is therefore consistent with the results from level
statistics. For $d=3$, a comparison between $\eta$ and the distribution of $P$ 
even allows to identify the ballistic regime for weak disorder, where the 
inverse participation ratio essentially remains constant.

In conclusion, we have shown that phase space properties represent a
new tool to describe disordered systems for arbitrary disorder
strength. The inverse participation ratio $P$ in phase space provides
information, which could not be obtained from the corresponding
quantities $P_x$ in real space and $P_k$ in momentum space. It is only
the first quantity which captures the appearance of a diffusive regime
in two and higher dimensions. Given our demonstration that the physics
in phase space is able to provide new insight into the dynamics of
disordered systems, these concepts likely will also make a valuable
contribution towards understanding the challenging problem of the 
combined effects of interaction and disorder in few-body systems.

\begin{acknowledgments}
This work was supported by the Son\-der\-for\-schungs\-be\-reich 484 of the
Deutsche Forschungsgemeinschaft. D.W. thanks the European Union for
financial support within the RTN program. The numerical calculations were 
carried out partly at the Leibniz-Rechenzentrum M{\"u}nchen.
\end{acknowledgments}

\end{document}